 \documentclass[final,1p,times]{elsarticle}
\usepackage{amssymb}
\usepackage{amsmath}
\usepackage[colorlinks=true,linkcolor=blue,citecolor=blue,urlcolor=blue]{hyperref}
\usepackage[normalem]{ulem}
\usepackage[table,xcdraw]{xcolor}

\begin{document}

\begin{frontmatter}



\title{Engineering of self-bending surface plasmon polaritons\\ through Hermite-Gaussian mode expansion}


\author{Javier Hernandez-Rueda$^*$, \'Angel S. Sanz and Rosario Mart{\'\i}nez-Herrero}
\cortext[cor1]{Corresponding author: fj.hernandez.rueda@ucm.es}

\affiliation{organization={Department of Optics, Faculty of Physical Sciences,
		Universidad Complutense de Madrid}, %Department and Organization
            addressline={\newline Pza.\ Ciencias 1}, 
            city={Madrid},
            postcode={28040}, 
            country={Spain}}

\begin{abstract}
Surface plasmon polaritons have received much attention over the last decades in photonics or nanotechnology due to their inherent high sensitivity to metal surface variations (e.g., presence of adsorbates or changes in the roughness).
It is thus expected that they will find promising major applications in widely cross-disciplinary areas, from material science to medicine.
Here we introduce a novel theoretical framework suitable for designing new types of structured paraxial surface plasmon beams and controlling their propagation.
More specifically, this method relies on a convenient Hermite--Gaussian mode expansion, which constitutes a complete basis set upon which new types of structured paraxial plasmon beams can be generated.
The family of beams generated in this way presents a rather peculiar feature: they exhibit local intensity maxima at different propagation distances, which enables the control over where to place the beam energy.
This, thus, opens up worthwhile pathways to manipulate light propagation along metal surfaces at the nanoscale.
As a proof-of-concept, we provide numerical evidence of the feasibility of the method by analyzing the propagation of Airy-based surface plasmon polaritons along an air--silver interface.
\end{abstract}

%\centering
%\includegraphics[width=0.75\columnwidth]{2CSDcase2}

%\item Paraxial partially coherent Airy beams are approached with a novel methodology.
%\item The methodology is based on a generalization of the concept of flux trajectory.
%\item It is aimed at investigating coherence and power content during propagation.
%\item The method stresses the role of phase relations against other density-based methods.
%\item It is applicable to any type of structured light regardless of its coherence degree.


\end{frontmatter}



%%%%%%%%%%%%%%%%%%%%%%%%%%%%%%%%%%%%%%%%%%%%%%%%%%%%%%%%%%%%%%%%%%%%%%%%
%%%%%%%%%%%%%%%%%%%%%%%%%%%%%%%%%%%%%%%%%%%%%%%%%%%%%%%%%%%%%%%%%%%%%%%%

\section{Introduction}
\label{sec1}

A surface plasmon polariton (SPP) is an electromagnetic excitation or wave that propagates along the interface between a dielectric medium and a metal.
Because of the dielectric--metal coupling, the amplitude of the SPP decreases exponentially fast with the propagation distance, while it remains confined within the vicinity of the\enlargethispage{-6pt} interface.
It is this appealing feature that makes SPPs highly sensitive to any property related to defects or changes present in the interface, such as the presence of adsorbates, the roughness of the metal surface, or variations in its chemical composition.
Due to this high sensitivity to surface properties, SPPs are highly appreciated in a myriad of applications
\cite{maradudin:PhysRep:2005,doi:10.1021/nl802044t,Manjavacas:09,Zhang:JPhysD:2012,xiangang:IEEEPhotJ:2012,anwar:DigCommunNet:2018}, which include guiding and control of light fields at subwavelength scales, super-enhanced Raman spectroscopy, solar cells, data storage, chemo- and bio-sensors, within the scope of what we could denote as classical light applications, although they extend further beyond to the quantum realm \cite{zhou:ProgQuantumElectron:2019}.
Hence, it is no wonder that the study of SPPs constitutes an active research area in optoelectronic technologies, biomedicine \cite{kneipp:JPCM:2002,saouli:MatToday:2020}, quantum communications \cite{altewischer:Naure:2002}, or metamaterials \cite{gric:JOpt:2017,gric:JIMTW:2018}.

The theoretical description of SPPs can be approached through rigorous many-body treatments focused on the electronic response of solids \cite{echenique:RepProgPhys:2006}.
Yet, it is more common resorting to Maxwell's equations, seeking for conditions that make the propagating electromagnetic field, assumed to be represented by a plane wave, to be p-polarized (i.e., perpendicular to the propagation along the interface in any direction).
Even if this level of description may be sufficient in many situations, it ignores any possible transversal structure.
Therefore, a complete characterization of the evolution and the confinement properties of SPPs requires a derivation beyond the single plane-wave approximation.
One thus wonders whether there are other optimal functional forms for SPP beams with clear advantages regarding the design of SPPs and the subsequent control of their propagation, in analogy to the so-called structured light beams \cite{forbes:NatPhotonics:2021}.

In 2007, the nondiffracting Airy beams, earlier conjectured by Berry and Balazs \cite{berry:AJP:1979}, were first experimentally produced with light by Christodoulides and coworkers \cite{christodoulides:OptLett:2007,christodoulides:PRL:2007}.
Soon they found remarkable applications in fields such as optical micromanipulation \cite{dholakia:NatPhoton:2008} or light-sheet microscopy \cite{dholakia:NatMethods:2014}.
Now, in connection to the question posed above, it is noteworthy that, shortly after their experimental implementation, the very same idea of structured light beams was proposed and produced in the case of SPPs, and also by the same group \cite{christodoulides:OL:2010,minovich:LasPhotRev:2014}.
These are the so-called Airy plasmons, which have been experimentally reproduced with alternative methods by different groups \cite{minovich:PRL:2011,li:PRL:2011,zhang:OL:2011,arie:LasPhotRev:2016,arie:ACSPhoton:2017}.
Airy plasmons exhibit the same properties as a usual Airy beam.
That is, they also undergo a transverse displacement that scales quadratically with the propagation distance, while the intensity profile remains shape-invariant during the propagation.
Nonetheless, there is an important difference between Airy plasmons and Airy beams: the intensity of the former exhibits an exponential decay as they propagate forward, as it also happens to any other type of SPP.
The interest in these nondiffracting beams thus relies on their noteworthy property of keeping, in homogeneous media, their main intensity maximum propagating along a precise parabolic trajectory without the need of controlling the process with the aid of nonlinear graded-index (GRIN) media \cite{Liu:OE:2019,raju-bk}.
Due to the interest generated by these diffraction-free plasmon beams, many other classes of self-bending SPPs have been subsequently proposed in the literature, such as the Pearcy SPPs \cite{hu:ResPhys:2021}, the Pearcy--Talbot SPPs \cite{ruan:OL:2024}, or the Olver SPPs \cite{chen:OL:2023}.

In this work, we introduce a new methodology aimed at designing and controlling the features of self-bending SPPs, which is based on a Hermite--Gaussian mode expansion of the electromagnetic excitation.
This approach is grounded on the theoretical framework set by Mart\'{\i}nez-Herrero and Manjavacas \cite{manjavacas:PRA:2016} to investigate SPP beams that propagate along the interface between a dielectric and a lossy metal under paraxial propagation conditions.
The approach here proposed is general and can be equally used to describe and/or infer propagation properties of any of the self-bending SPPs mentioned above, yet here we constrain ourselves to the particular case of Airy-type plasmons with a finite energy content.
Accordingly, the work has been organized as follows.
In \hyperref[sec2]{Section~\ref{sec2}}, we introduce and discuss the decomposition method in terms of Hermite--Gaussian modes applied to self-bending SPPs with a finite energy content.
The application of this methodology to the analysis of the propagation properties displayed by Airy-based self-bending SPPs, as a particular class of self-bending SPPs, is accounted for in \hyperref[sec3]{Section~\ref{sec3}}.
Finally, the main conclusions arising from the method here introduced are summarized in \hyperref[sec4]{Section~\ref{sec4}}.

%%%%%%%%%%%%%%%%%%%%%%%%%%%%%%%%%%%%%%%%%%%%%%%%%%%%%%%%%%%%%%%%%%%%%%%%
%%%%%%%%%%%%%%%%%%%%%%%%%%%%%%%%%%%%%%%%%%%%%%%%%%%%%%%%%%%%%%%%%%%%%%%%

\section{Hermite-Gaussian mode decomposition for finite energy self-bending SPPs}
\label{sec2}

Following the procedure introduced by Mart\'{\i}nez-Herrero and Manjavacas \cite{manjavacas:PRA:2016} in 2016 to describe the propagation of Gaussian SPPs in terms of modes, below we proceed in the same way considering the case of any general paraxial self-bending SPP with a structure modulated by the phase of the beam's angular spectrum.
Moreover, we also consider conditions of finite energy, thus implicitly assuming realistic experimental realization scenarios.
Accordingly, in the paraxial approximation [see \hyperref[eq7]{Eq.~(\ref{eq7})}], we can write the general expression for a finite energy self-bending SPP at a dielectric--metal interface along the $x-z$ plane as
\begin{align}
 \psi_0(x,z) = \int e^{i\varphi(u)} \xi(u) e^{i|k_{\text{spp}}|ux} e^{-ik^{*}_{\text{spp}} zu^{2}/2} du, 
 \label{eq20}
\end{align}
where $k_{\textrm{spp}}$ is the SPP wave number and $\varphi(u)$ is a real-valued polynomial determining the type of self-bending SPP.
Concerning the phase function $\varphi(u)$, it is worth mentioning that for $\varphi(u) = u^3/3$ we have an Airy plasmon, while $\varphi(u) = u^4$ gives rise a Pearcy plasmon.
Because of the SPP finite energy content, its transversal intensity, \hyperref[eq9]{Eq.~(\ref{eq9})}, is also finite, which implies
\begin{align}
 \int |\xi(u)|^2 du < \infty .
 \label{eq21}
\end{align}

To describe $\xi(u)$ and generate a suitable basis set that will allow us to analyze and characterize the propagation of the self-bending SPPs described by \hyperref[eq20]{Eq.~(\ref{eq20})}, we introduce the Hermite--Gaussian modes,
\begin{align}
 \phi_n(u) = a_n H_n (\alpha u) e^{-\alpha^2u^2/2}, 
 \label{eq22}
\end{align}
where $H_n$ is the $n$th-order Hermite polynomial, $\alpha$ is a dimensionless constant, and
\begin{align}
 a_n^2 = \frac{\alpha}{2^n n!\sqrt{\pi}} .
 \label{eq23}
\end{align}
The $\phi_n$ modes constitute a complete set of orthonormal functions determined by a single parameter, namely, $\alpha$.
The square integrable function $\xi(u)$ can be recast as a linear superposition of these basis functions, as\enlargethispage{-12.5pt}
\begin{align}
 \xi(u) = \sum c_n \phi_n(u), 
 \label{eq24}
\end{align}
with
\begin{align}
 c_n = \int \phi_n (u) \xi (u) du .
 \label{eq25}
\end{align}
If \hyperref[eq24]{Eq.~(\ref{eq24})} is substituted into \hyperref[eq20]{Eq.~(\ref{eq20})}, then we obtain a mode-based description for the field amplitude of any arbitrary paraxial self-bending SPP,
\begin{align}
 \psi_0(x,z) = \sum c_n \phi_{n0} (x,z), 
 \label{eq26}
\end{align}
where each mode has the functional form
\begin{align}
 \phi_{n0}(x,z) = \int e^{i\varphi (u)} \phi_n(u) e^{i|k_{\textrm{spp}}|xu - ik^*_{\textrm{spp}} zu^2/2} du .
 \label{eq27}
\end{align}
However, note that, because the coefficients $c_n$ depend on the particular choice of $\alpha$ (each value of $\alpha$ defines a family of modes), there is not a unique manner to choose the basis set that will be used to expand a given self-bending SPP.
Here, in particular, we have chosen $\alpha$ in a way that it results in an expansion that best fits the actual beam with the lowest number of terms, while it still preserves the essential features of the corresponding self-bending SPP.
From now on, we shall focus on the propagation of those single mode components rather than on the full parent beam, \hyperref[eq20]{Eq.~(\ref{eq20})}, as they define a family with interesting properties. Below we discuss some global properties, which are valid for any self-bending SPP, regardless of its functional form, and later will consider an application to paraxial Airy-type SPPs.
Under realistic experimental conditions, to the best of our knowledge, this type of self-bending SPP beam decomposition could be experimentally realized by employing established structured-light strategies for controlled light--surface coupling. Programmable spatial light modulators could be used for precise imprinting of the Hermite--Gaussian amplitude and phase profiles onto an incident beam before coupling it via holographic gratings; see, for instance, Dolev et~al.~\cite{dolev2012surface}. Alternatively, diffractive optical elements directly engraved in the metal layer can be employed to encode the required decomposition at the metal--dielectric interface, where the resulting SPP generation can be measured by scanning near-field optical microscopy \cite{klein2012controlling,minovich2014airy}.

From \hyperref[eq20]{Eq.~(\ref{eq20})}, we note that the angular spectral component of the propagated field, $\psi(x,z)$, can be recast as $\tilde{\psi}_0(u) = e^{i\varphi(u)} \xi(u)$ (see \hyperref[appB]{\ref{appB}}).
Now, global properties, such as the transversal intensity or its variation along the $z$-coordinate, do not depend on the phase of $\tilde{\psi}_0(u)$, i.e., $\varphi(u)$.
The same is expected for the quantities associated with the modes $\phi_{n0}(x,z)$.
This is precisely what we find when these two quantities are computed:
\begin{align}
\bar{I}_{n0}(z) & =  \frac{2\pi a_n^2}{|k_{\textrm{spp}}|} \int H_n^2(\alpha u) e^{- [\alpha^2 + k''_{\textrm{spp}}z] u^2} du, 
\label{eq28} \\[12pt]
\bar{I}'_{n0}(z) & =  - \frac{2\pi a_n^2~k''_{\textrm{spp}}}{|k_{\textrm{spp}}|} \int u^2 H_n^2(\alpha u) e^{- [\alpha^2 + k''_{\textrm{spp}}z] u^2} du,   
\label{eq29}
\end{align}
where $\bar{I}_{n0}(z) = \int |\phi_{n0}(x,z)|^2 dx$ and $\bar{I}'_{n0}(z)$ is the spatial derivative defined by \hyperref[eq11]{Eq.~(\ref{eq11})}.
As a consequence, although the spatial structure of the self-bending SPP depends on $\varphi(u)$, both the transversal intensity for these SPPs and its variation along the $z$-direction are going to exhibit the same functional form regardless of the type of SPP considered.
The only difference relies on the specific value assigned to $\alpha$, which may differ from one type of self-bending SPP to another one, as it is chosen taking into account the above-mentioned best-fit criterion.

\begin{figure}[!t]
\centering
\includegraphics[width=0.5\columnwidth]{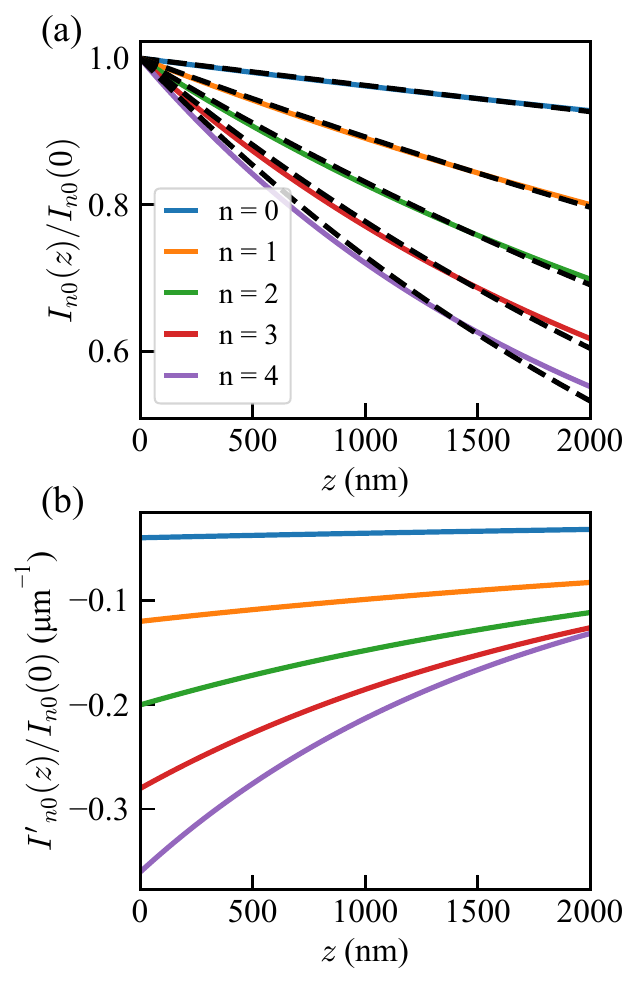}
\caption{$\bar{I}_{n0}(z)$ (a) and $\bar{I}'_{n0}(z)$ (b) as a function of the $z$-distance for the five lowest-order modes.  The color code used is: blue for $n=0$, orange for $n=1$, green for $n=2$, red for $n=3$, and purple for $n=4$. Note that the $z$-range covers up to 2~$\mu$m, quite below the propagation length, $L_{\rm spp} \approx 60$~$\mu$m.}
\label{fig1}
\end{figure}

In \hyperref[fig1]{Fig.~\ref{fig1}}, we show the dependence on $z$ for both $\bar{I}_{n0}(z)$ (a) and $\bar{I}'_{n0}(z)$ (b) for the five lowest-order modes for an air--silver interface and $\alpha = 0.316$.
We have considered silver, because of its low loss at optical frequencies among all metals.
Yet there are discrepancies in the value of its relative permittivity \cite{johnson:PRB:1972,yang:PRB:2015,mcpeak:ACSPhoton:2015,jiang:SciRep:2016,ferrera:PhysRevMater:2019}, which arise from the different ways to prepare samples in the laboratory \cite{Wu:AdvMat:2014,Wu:AdvMat:2014corr}.
Bearing this in mind, to produce the results shown here, we have considered $\varepsilon_m = -18.3132 + 0.49806 i$, at a 633-nm wavelength (extracted from data reported by Johnson and Christy \cite{johnson:PRB:1972}, widely used within the plasmonics community).
Thus, in panel~\ref{fig1}(a), we note a monotonic decay, with a faster fall as $n$ increases, from $n=0$ (blue solid line) to $n=4$ (purple solid line).
In this particular case, the value of the propagation length \eqref{Lspp} is $L_{\textrm{spp}} \approx 60$~{\textmu}m, which renders a falloff much slower even for the lowest orders.
To quantify this trend, we have fit the curves to a decaying exponential, $e^{-\gamma z}$, which has rendered the following decay lengths, $\gamma^{-1}$: 26.42~{\textmu}m for $n = 0$, 8.8~{\textmu}m for $n = 1$, 5.4~{\textmu}m for $n = 2$, 4.0~{\textmu}m for $n = 3$, 3.2~{\textmu}m for $n = 4$.
As it can be noticed, all these characteristic values are quite below the propagation length $L_{\textrm{spp}}$, which justifies our analysis of the propagation in terms of only $\psi_0(x,z)$, excluding the exponential prefactor of the full field amplitude, $e^{- {\textrm{Im}}(k_{\textrm{spp}}) z}$ (see \hyperref[appA]{~\ref{appA}}).
Now, with respect to the curves displayed in panel~\ref{fig1}(b), they show that the decay of the respective $I_{n0}(z)$ functions is not constant, i.e., strictly speaking, it is not a purely single-parameter exponential, although it is close to it.
What we infer instead is that there is a fast fall-off at short distances, while the falling rate relaxes for larger values of $z$.
Also note that, as $n$ increases, this effect becomes more apparent (compare the case for $n=0$, almost flat in the $z$-range shown, with that for $n=4$, for instance).

Additionally, we may wish to determine global phase information, which happens to also be independent of the phase $\varphi(u)$.
To this end, if $\phi_{n0}(x,z)$ is recast in polar form [see \hyperref[eq14]{Eq.~(\ref{eq14})}], as
\begin{align}\label{ieq1}
 \phi_{n0}(x,z) = A_{n0}(x,z) e^{i S_{n0}(x,z)}, 
\end{align}
the expression for the average phase variations along the $z$ direction for each mode reads as
\begin{align}\label{ieq2}
 \left\langle \frac{\partial S_{n0}}{\partial z} \right\rangle\! (z) =
	\int I_{n0}(x,z) \frac{\partial S_{n0}(x,z)}{\partial z} dx, 
\end{align}
which renders
\begin{align}
 \left\langle \frac{\partial S_{n0}}{\partial z} \right\rangle\! (z) =  - \frac{\pi a_n^2~k'_{\textrm{spp}}}{|k_{\textrm{spp}}|} \! \int u^2 H_n^2(\alpha u) e^{- [\alpha^2 + k''_{\textrm{spp}} z] u^2} du
 =  \frac{k'_{\textrm{spp}}}{2k''_{\textrm{spp}}} \bar{I}'_{n0}(z), 
 \label{eq30}
\end{align}
in agreement with \hyperref[eq15]{Eq.~(\ref{eq15})}.
This quantity divided by $\bar{I}_{n0}(0)$ is proportional to the results presented in \hyperref[fig1]{Fig.~\ref{fig1}}(b) for the five modes considered. Therefore, the results provide us with a different physical perspective on these outcomes: while the transversal intensity undergoes a progressive decay, faster as $n$ increases, the trend shown by the average phase variation along the $z$-direction indicates a gradual approach to a constant value.

%%%%%%%%%%%%%%%%%%%%%%%%%%%%%%%%%%%%%%%%%%%%%%%%%%%%%%%%%%%%%%%%%%%%%%%%

\section{Paraxial finite-energy Airy plasmon propagation}
\label{sec3}

We now consider the propagation of paraxial Airy and Airy-type plasmons.
This particular case fits the functional form of $\varphi(u) = u^3/3$ in \hyperref[eq20]{Eq.~(\ref{eq20})}, where the modes given by \hyperref[eq27]{Eq.~(\ref{eq27})} read as
\begin{align}
 \phi_{n0}(x,z) & =  a_n \int e^{iu^3/3} H_n(\alpha u) e^{-\alpha^2u^2/2} e^{i|k_{\textrm{spp}}| xu - ik^*_{\textrm{spp}}zu^2/2} du .
 \label{eq31}
\end{align}
These modes can be recast in terms of the Airy function and its higher-order derivatives, as\enlargethispage{-5.7pt}
\begin{align}
 \phi_{n0}(x,z) = a_n e^{2\beta_z^2/3} e^{|k_{\textrm{spp}}| \beta_z x} 
 \sum_{s=0}^n \frac{n!}{s!(n-s)!} {\left(\frac{2\alpha}{i|k_{\textrm{spp}}|} \right)}^{n-s} H_s (-i\alpha \beta_z)\ \frac{\partial^{n-s}}{\partial x^{n-s}} Ai (|k_{\textrm{spp}}| x + \beta_z^2), 
 \label{eq32}
\end{align}
where
\begin{align}
 \beta_z := \frac{\alpha^2 + ik_{\textrm{spp}}^* z}{2} .
 \label{eq33}
\end{align}

\begin{figure}[!t]
\centering
\includegraphics[width=0.95\columnwidth]{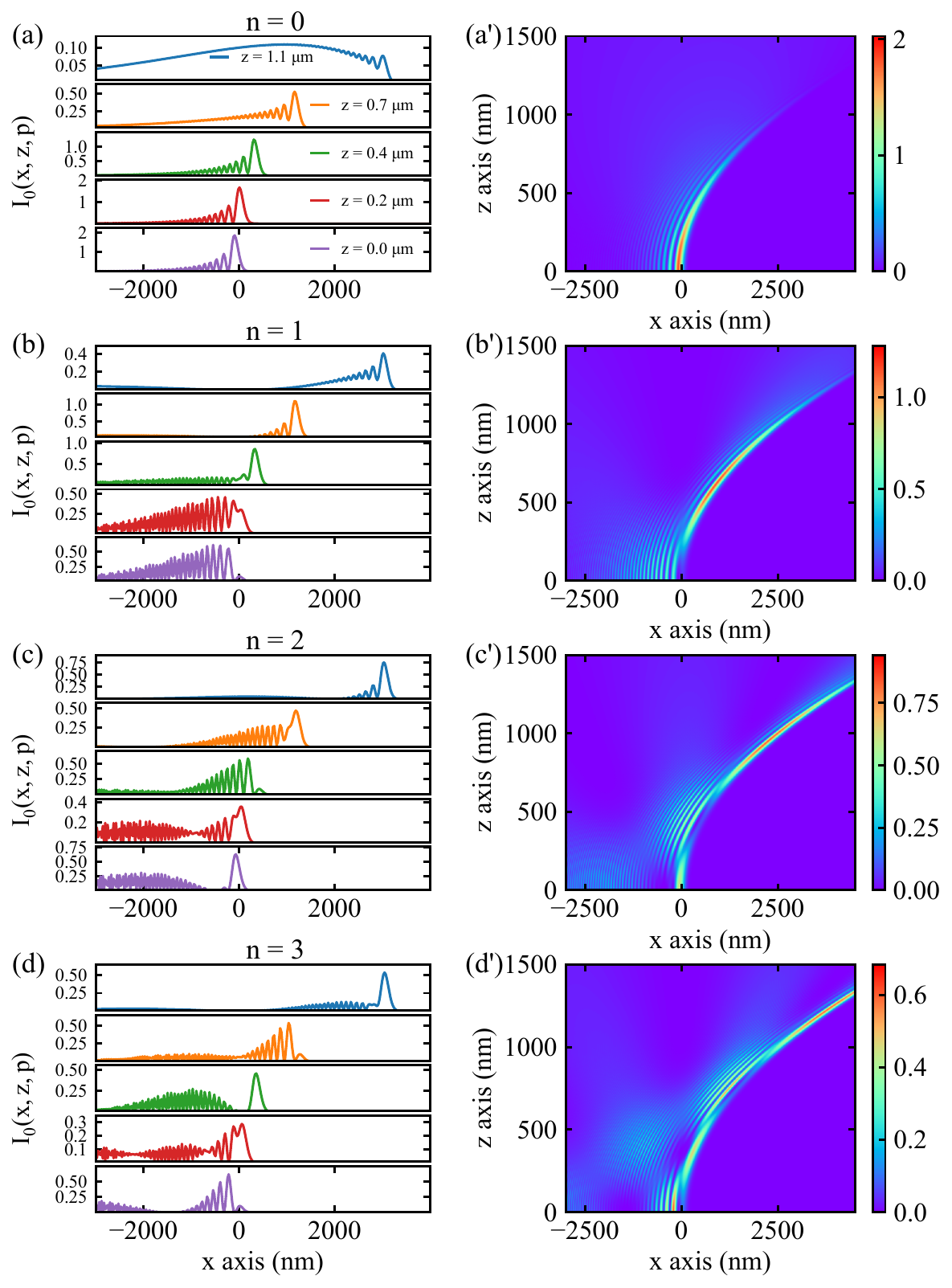}
\caption{Snapshots (left column) and density plots (right column) illustrating the propagation of the Airy-type plasmons with lowest order $n$: (a/a') $n=0$, (b/b') $n=1$, (c/c') $n=2$, and (d/d') $n=3$. In the snapshots, we have used the following line colors: purple for $z=0$, red for $z=0.2$~$\mu$m, green for $z=0.4$~$\mu$m, orange for $z=0.7$~$\mu$m, and blue for $z=1.1$~$\mu$m.}
\label{fig2}
\end{figure}

\hyperref[fig2]{Figure~\ref{fig2}} shows the propagation of the four lowest-order Airy-type plasmons (i.e., from $n=0$ to $n=3$).
In the left column, in each panel, from (a) to (d), we observe a series of snapshots of the intensity $I_{n0}(x,z)$, \nobreak{}covering propagation distances from $z = 0$ to $z = 1.1$~{\textmu}m.
By inspecting these plots, we note that the lowest order, $n=0$, illustrates an evident diffusion of the intensity along the $x$-direction, while for higher orders we observe the opposite behavior, that is, a higher localization of the intensity within a rather narrow region.
For free finite-energy Airy beams, it is known \cite{sanz:PRA:2022,sanz:JOSAA:2022,sanz:OE:2024} that there is a back flow of energy, which leads to the suppression of the defining traits of an Airy beam.
In contrast, this does not happen in ideal Airy beams, because of the continuous push forward received by the rearmost part of the intensity distribution, i.e., the infinite tail.  
Here, high-order Airy-type SPPs show how some energy is released backwards at some values of $z$, which generates a rather long tail, although with a relatively low intensity in all its extension.
Such energy release generates a sort of recoil effect, which makes the front part to preserve its shape while it propagates forwards.

Unlike the behavior displayed by the standard Airy plasmon, shown in \hyperref[fig2]{Fig.~\ref{fig2}}(a), in the propagation of the Airy-type plasmons displayed in panels (b)--(d) we note the splitting of the main maximum of the \nobreak{}intensity distribution into two at some specific values of the propagation coordinate.
If instead of the leftmost snapshot profiles, we focus on the density plots of the right-column panels, which show the respective density plots illustrating the full propagation from $z=0$ to $z=1.5$~{\textmu}m, we can better appreciate that these splittings are a consequence of a redistribution of the intensity between two adjacent seemingly parabolic paths.
As it can be noticed, the transition between these two pathways is in correspondence with the order of the mode, that is, no crossover for $n=0$, one crossover for $n=1$ (at $z \approx 250$~nm), a double crossover for $n=2$ (first at $z \approx 250$~nm and then at $z \approx 700$~nm), etc.
Furthermore, apart from this splitting, we also observe another interesting phenomenon.
In panel (a'), we only find a gradual suppression and dispersion of the Airy SPP, similar to the behavior exhibited by a free finite-energy Airy beam \cite{sanz:PRA:2022,sanz:JOSAA:2022}, which gets gradually blurred because of the back flow of energy enabled by the finite nature of the tail
[this is, precisely, what we observe in panel (a')].
However, in panels (b') to (d'), the situation is different, as they present a more interesting and unexpected behavior: the transition between the two pathways gives rise to a displacement of the region with a maximum intensity to larger $z$-distances.

These two behaviors, splitting of the main intensity maximum and shift of the maximum intensity region to larger $z$-values, arise from the same origin, namely, the fact that all these self-bending SPPs are describable in terms of a combination of an Airy function and its first derivative.
Note here that, according to \hyperref[eq32]{Eq.~(\ref{eq32})}, the propagation of these engineered Airy-type plasmons can be interpreted in terms of a superposition (interference) of high-order derivatives of the Airy-type plasmons.
Now, all these higher-order derivatives can be recast in terms of only the Airy function itself and its first derivative \cite{brychkov:ITSF:2012,laurenzi:arxiv:2017,razueva:SIGMA:2018}.
Given that their argument is complex valued, there is going to be a $z$-dependent phase shift between these two functions, such that, as $z$ increases, interference traits will appear.
This can readily be seen if we inspect the explicit form of the modes represented in \hyperref[fig2]{Fig.~\ref{fig2}}.
Neglecting the common exponential prefactors, we find
\begin{align}
 \phi_{00}(x,z) & \sim  Ai (x,z), 
 \label{eq41} \\[6pt]
 \phi_{10}(x,z) & \sim  -2i\alpha \left[ \beta_z Ai (x,z) + A_i' (x,z) \right], 
 \label{eq42} \\[6pt]
 \phi_{20}(x,z) & \sim  -2 \left[ (1 + 2\alpha^2 |k_{\textrm{spp}}|x + 4 \alpha^2\beta_z^2) Ai (x,z) + 4 \beta_z A_i' (x,z) \right], 
 \label{eq43} \\[6pt]
 \phi_{30}(x,z) & \sim  4i\alpha \left[ (3 + 2 \alpha^2 |k_{\textrm{spp}}| x + 8\alpha^2\beta_z^2) A_i' (x,z) \right. \nonumber \\[6pt]
 & \quad + (2\alpha^2 + \left. 3\beta_z + 6\alpha^2\beta_z |k_{\textrm{spp}}| x + 8\alpha^2 \beta_z^3) Ai (x,z) \right] ,
 \label{eq44}
\end{align}
where the explicit argument of the Airy function and its derivative, $A_i'$, is $|k_{\textrm{spp}}| x + \beta_z^2$.
According to this set of equations, for $n>0$, the Airy function and its first derivative are both affected by a $z$-dependent factor through the $\beta_z$-function and its powers.
For odd $n$, $Ai(x,z)$ and $Ai'(x,z)$ are multiplied by odd and even powers of $\beta_z$, respectively, whereas it is just the opposite for even $n$.
To understand the role of $\beta_z$ here, note that, for $n=0$, we have a pure Airy function, which is basically what we observe in \hyperref[fig2]{Fig.~\ref{fig2}}(a).
However, in \hyperref[fig2]{Fig.~\ref{fig2}}(b), for $n=1$, at $z=0$, the contribution of the Airy function is proportional to $\alpha^2$, while its derivative contributes at full.
Therefore, at $z=0$, the main contribution will arise from $Ai'(x,z)$.
However, the contribution from the Airy function increases linearly with $z$, according to \hyperref[eq42]{Eq.~(\ref{eq42})}.
So, as soon as a threshold value is overcome, $Ai(x,z)$ is going to acquire the leading role, which will remain for the rest of the propagation.
This is precisely what we observe in \hyperref[fig2]{Fig.~\ref{fig2}}(e), where there is a crossover around $z \approx 250$~nm.

For $n=2$, we find the opposite situation at $z=0$: the Airy function constitutes the leading term, while its derivative appears multiplied by a prefactor $\alpha^2$.
As a result, as seen in \hyperref[fig2]{Fig.~\ref{fig2}}(c), we observe a maximum at the position of the maximum for the usual Airy function, although it is followed by a different structure of secondary maxima, which is due to the contribution coming from $Ai'(x,z)$.
Besides, we also note that, according to \hyperref[eq43]{Eq.~(\ref{eq43})}, a linear term in $z$ affects both $Ai(x,z)$ and $Ai'(x,z)$, although in the case of the former it is multiplied by $\alpha^2$.
Therefore, at intermediate $z$ values, we expect a prevalence of $Ai'(x,z)$.
This is, actually, what we observe in \hyperref[fig2]{Fig.~\ref{fig2}}(c'), at $z \approx 250$~nm, where the back crossover indicates the domain of $Ai'(x,z)$ over $Ai(x,z)$.
Nonetheless, given that $Ai(x,z)$ is multiplied by $\beta_z^2$, at longer $z$-distances, the corresponding quadratic term in $z$ is expected to gain more relevance, and hence there will be another crossover forwards, so that $Ai(x,z)$ will recover the leading role.
From \hyperref[fig2]{Fig.~\ref{fig2}}(c'), we note that this happens, effectively, at $z \approx 700$~nm.
An analogous trend also serves to explain the behavior for $n=3$, although there will be three crossovers, since a third power of $\beta_z$ is involved.
In general, and before the SPP dissipates, one thus expects as many interferential crossovers as the highest power of $\beta_z$ or, equivalently, the degree of the mode $\phi_{n0}$, i.e., $n$.\enlargethispage{-4.3pt}

It is also worth noting that, if we compare the results of \hyperref[fig2]{Fig.~\ref{fig2}} with those displayed in \hyperref[fig1]{Fig.~\ref{fig1}}(a), we reach a rather puzzling situation: how is it possible that the relative transversal energy decreases faster as $n$ increases [see \hyperref[fig1]{Fig.~\ref{fig1}}(a)] while the intensity maxima are larger in the higher orders?
To clarify this counter-intuitive situation, let us consider the energy content at $z = 1.1$~{\textmu}m.
We note that, in the lowest order, the intensity distributes over a wide region and with a height nearly of the same order.
This ensures a relatively high power content and hence a weak energy loss [about $5~\%$, by inspecting \hyperref[fig1]{Fig.~\ref{fig1}}(a)].
In contrast, as $n$ increases, we observe a highly localized intensity distribution followed by a low tail, which, when integrated over the whole $x$-range, gives rise to higher losses [up to $30~\%$ in the case of $n=4$; see \hyperref[fig1]{Fig.~\ref{fig1}}(a)].

\begin{figure}[!t]
\centering
\includegraphics[width=0.6\columnwidth]{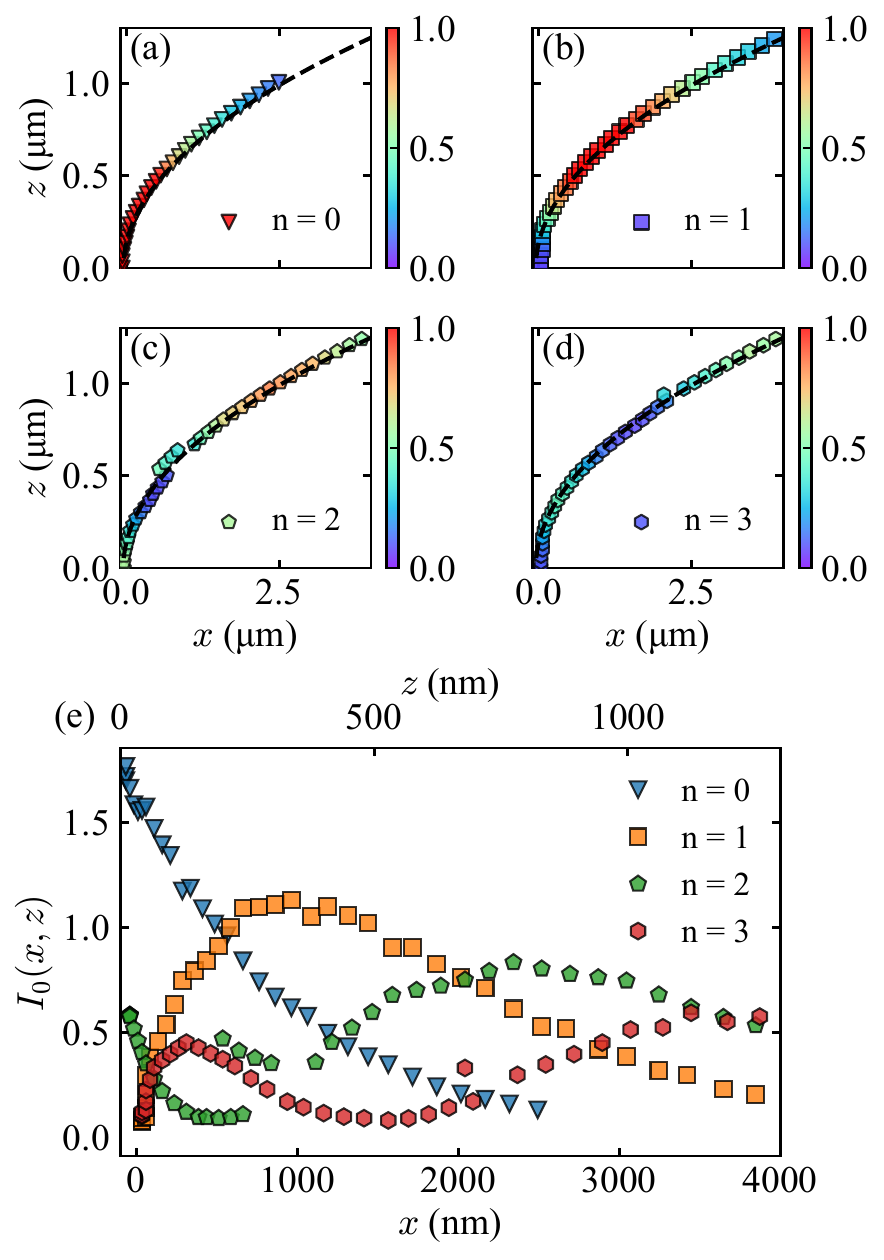}
\caption{Panels (a) to (d) show the nearly parabolic trajectory (black dashed line) followed by the leading maximum of the Airy-type plasmons considered in Fig.~\ref{fig2}: (a) $n=0$, (b) $n=1$, (c) $n=2$, and (d) $n=3$. (e) Intensity along the trajectory followed by the leading maximum for the four cases shown in the above panels.}
\label{fig3}
\end{figure}

Finally, in order to set optimal control scenarios based on the local value of the intensity, it seems reasonable to determine the precise $x-z$ location of the intensity maxima as the self-bending SPPs propagate.
Of course, the maximum value of the intensity is going to be located along the direction followed by the leading maximum of the Airy-type plasmon.
\hyperref[fig3]{Fig.~\ref{fig3}}(a)--(d) show that, regardless of the order, all leading maxima propagate, in a fair approximation, along a parabolic trajectory, basically the same for the four orders displayed.
For the orders $n=2$ and $n=3$ [\hyperref[fig3]{Fig.~\ref{fig3}}(c) and (d), respectively], there is an apparent deviation, but this is only due to the presence of the above-mentioned crossovers; what really matters here, in any case, is the general trend.
If we now focus on the intensity along these trajectories, we obtain an interesting picture, as \hyperref[fig3]{Fig.~\ref{fig3}}(e) reveals. The graph illustrates that, as the order $n$ increases, the relative maximum of the intensity undergoes a displacement towards higher values of both $z$ and $x$ (see upper and bottom x-axes).
Accordingly, some Airy-type plasmons turn out to be more convenient than others depending on the distance with respect to the input plane, where we wish to allocate the maximum power.
Note that these intensity distributions follow a neat pattern: as $n$ increases, not only the maximum moves towards higher values of $z$, but the number of relative maxima is given by $n$. These specific propagation features can be leveraged to generate intense light hotspots and tailor their spatial features, for instance by overlapping pairs of Airy SPPs with the same order $n$ \cite{martinezherrero2025}. Moreover, we also observe that, while for even powers of $n$ the intensity always falls from a finite value, for odd $n$  the intensity starts from zero.

%%%%%%%%%%%%%%%%%%%%%%%%%%%%%%%%%%%%%%%%%%%%%%%%%%%%%%%%%%%%%%%%%%%%%%%%
%%%%%%%%%%%%%%%%%%%%%%%%%%%%%%%%%%%%%%%%%%%%%%%%%%%%%%%%%%%%%%%%%%%%%%%%

\section{Final remarks}
\label{sec4}

Currently, the growing interest of the photonics community in SPPs drives the increasing need to develop methodologies that assist in designing novel SPP types and controlling their energy allocation during propagation. Our work tackles this problem by means of a convenient Hermite--Gaussian mode expansion that fully describes self-bending paraxial SPPs. We have employed this mode expansion to investigate the propagation of Airy plasmon-based beams at an air--silver interface, although our method is not solely circumscribed to this specific type of light beam.

For Airy-type plasmons, a prototype of self-bending SPPs, we have shown that all modes with $n>0$ allow us to allocate a relatively large amount of energy at considerably large distances away from the input plane (i.e., of the order of a few microns, with the physical conditions here considered). The ability to manipulate the precise location of the SPPs energy directly points out to applications that require some sort of focusing. Thus, our method holds great potential for applications that require highly localized light at the nanoscale, such as optical trapping and pulling techniques exploited by plasmonic-based tweezers.



%%%%%%%%%%%%%%%%%%%%%%%%%%%%%%%%%%%%%%%%%%%%%%%%%%%%%%%%%%%%%%%%%%%%%%%%
%%%%%%%%%%%%%%%%%%%%%%%%%%%%%%%%%%%%%%%%%%%%%%%%%%%%%%%%%%%%%%%%%%%%%%%%

\appendix

\section{General theoretical framework}
\label{appA}

In the description of the SPP propagation, we have considered that the $X\!Z$-plane $y=0$ coincides with the physical interface, while the $y$-axis points along the perpendicular direction, from the metal (in the $y < 0$ region) to the dielectric (in the region $y > 0$).
The dielectric medium is characterized by a relative permittivity $\varepsilon_d$, and the metal by $\varepsilon_m = \varepsilon'_m + i \varepsilon''_m$, with $\varepsilon'_m < 0$ and  $\varepsilon''_m > 0$, such that $\varepsilon_d + \varepsilon'_m < 0$.

It can be shown \cite{Archambault:PRB:2009,teperik:OE:2009,friberg:OL:2013,manjavacas:OE:2015,manjavacas:PRA:2016} that the electric field of a monochromatic SPP, with frequency $\omega$, in the dielectric medium can be written as
\begin{equation}
	{\bf E}_d (x,y,z) = {\bf E}_{d0} (x,z)  e^{ik_{dy} y} = \left( \begin{array}{c}
		\displaystyle i\frac{\partial \psi (x,z)}{\partial x} \\ \displaystyle \frac{\varepsilon_m}{\varepsilon_d} k_{dy} \psi(x,z) \\ \displaystyle i\frac{\partial \psi(x,z)}{\partial z} \end{array} \right) e^{ik_{dy} y} .
	\label{eq1}
\end{equation}
%
%\begin{eqnarray}
%	{\bf E}_d (x,y,z) & = & {\bf E}_{d0} (x,z)  e^{ik_{dy} y} \nonumber \\
%	& = & \left[ i\frac{\partial \psi (x,z)}{\partial x}, \frac{\varepsilon_m}{\varepsilon_d} k_{dy} \psi(x,z), i\frac{\partial \psi(x,z)}{\partial z} \right] e^{ik_{dy} y} . \nonumber \\ & &
%	\label{eq1}
%\end{eqnarray}
%
In this expression, $\psi(x,z)$ is a scalar field satisfying the surface Helmholtz equation,
\begin{equation}
	\frac{\partial^2 \psi(x,z)}{\partial x^2} + \frac{\partial^2 \psi(x,z)}{\partial z^2} + k_{\rm SPP}^2 \psi(x,z) = 0 ,
	\label{eq1b}
\end{equation}
and
\begin{equation}
	k_{dy} = k_{\rm SPP} \sqrt{\frac{\varepsilon_d}{\varepsilon_m}} ,
	\label{eq4}
\end{equation}
where
\begin{equation}
	k_{\rm SPP} = k'_{\rm SPP} + i k''_{\rm SPP} = k_0 \sqrt{\frac{\varepsilon_m \varepsilon_d}{\varepsilon_m + \varepsilon_d}} .
	\label{eq3}
\end{equation}
is the SPP wave number, with $k_0 = \omega/c$.
According to Eq.~\eqref{eq1}, in principle, the SPP propagation along the metal-dielectric interface depends on polarization and, therefore, it has a vector nature.
However, because the ratio between the $x$ and $z$ field components and the $y$-component is of the order of $1/\sqrt{|\varepsilon_m|}$, the contribution of the former can be neglected.
This allows us to consider \cite{manjavacas:OL:2017}, in a good approximation, that the value of the intensity distribution will be proportional to $|\psi(x,z)|^2$.

By using the angular spectrum method \cite{MartinezHerrero:OE:2008,manjavacas:OE:2015}, one obtains a solution to Eq.~\eqref{eq1b} with the following functional form
\begin{align}
	\psi(x,z) = \int \tilde{\psi}(u,z) e^{i |k_{\rm SPP}| ux} du = \int \tilde{\psi}_0(u) e^{i |k_{\rm SPP}| ux + ik_z(u) z} du ,
	\label{eq2}
\end{align}
for $z \ge 0$, where $k_z(u) = \sqrt{k_{\rm SPP}^2 - |k_{\rm SPP}|^2 u^2}$ and, to simplify notation, we have defined $\tilde{\psi}_0(u) := \tilde{\psi}(u,0)$.
If now assume that the SPP energy mainly propagates along the $z$-direction, we can consider the paraxial form of Eq.~\eqref{eq2},
\begin{equation}
	\psi(x, z) = \psi_0 (x, z) e^{i k_{\rm SPP} z} ,
	\label{eq5}
\end{equation}
where $\psi_0(x,z)$ satisfies Helmholtz's paraxial equation,
\begin{equation}
	\frac{\partial^2 \psi_0(x,z)}{\partial x^2} + 2 i k_{\rm SPP} \frac{\partial \psi_0(x,z)}{\partial z} = 0 .
	\label{eq6}
\end{equation}
The solution \eqref{eq5} can be determined by means of the angular spectrum method from Eq.~\eqref{eq6}, but also directly from the general solution \eqref{eq2}, assuming $u^2 \ll 1$ in the expression $k_z(u)$.
Accordingly, $k_z(u) \approx k_{\rm SPP} - |k_{\rm SPP}|^2 u^2/2k_{\rm SPP} = k_{\rm SPP} - k_{\rm SPP}^* u^2/2$, such that the first contribution gives rise to the exponential factor in \eqref{eq5}, and the second one describes the propagation along the $z$-direction inside the integral.
We thus obtain
\begin{equation}
	\psi_0(x,z) = \int_{-\infty}^\infty \tilde{\psi}_0(u) e^{i |k_{\rm SPP}| xu - ik_{\rm SPP}^* z u^2/2} du ,
	\label{eq7}
\end{equation}
which can be interpreted as a superposition of inhomogeneous two-dimensional (scalar) fields.
These fields contribute with a weight $|\tilde{\psi}_0(u)|$, although all of them decay with the same rate, $e^{- k''_{\rm SPP} z u^2/2}$, as the plasmon propagates ahead along the $z$-direction.
Note that this intrinsic decay only affects $\psi_0(x,z)$.
However, in Eq.~\eqref{eq5}, there is another contribution coming from the exponential term, which also induces an exponential attenuation of the SPP intensity distribution along the $z$-direction, $e^{-2 k''_{\rm SPP} z}$.
This introduces the propagation length \cite{raether-bk},
\begin{equation}
	L_{\rm SPP} = \frac{1}{2 k''_{\rm SPP}} \approx \frac{\lambda_0}{2\pi} \left( \frac{\varepsilon_d + \varepsilon'_m}{\varepsilon_d \varepsilon'_m} \right)^{3/2} \frac{{\varepsilon'_m}^2}{\varepsilon''_m},
 \label{Lspp}
\end{equation}
where we assume $|\varepsilon'_m| \gg \varepsilon_d, \varepsilon''_m$, and which provides us with information about the distance beyond which the SPP intensity has fallen to $e^{-1}$ its value at $z=0$ when the SPP propagates along a smooth surface.
Thus, in order to investigate experimentally the properties of the field $\psi_0(x,z)$, $L_{\rm SPP}$ has to be chosen in such a way that the decay of $\psi(x,z)$ takes place at $z$-distances much longer than those involved by the decaying term $e^{- k''_{\rm SPP} z u^2/2}$.

%%%%%%%%%%%%%%%%%%%%%%%%%%%%%%%%%%%%%%%%%%%%%%%%%%%%%%%%%%%%%%%%%%%%%%%%

\section{Transversal intensity and phase of a self-bending SPP}
\label{appB}

Bearing in mind the above assumption on $L_{\rm SPP}$, from now on we focus on $\psi_0(x,z)$.
To better understand the propagation properties of the SPP, we recast $\psi_0(x,z)$ in polar form, as
\begin{equation}
	\psi_0(x,z) = A(x,z) e^{i S(x,z)} ,
	\label{eq14}
\end{equation}
where $A(x,z)$ is the field amplitude, such that $I_0(x,z) = A^2(x,z) = |\psi_0(x,z)|^2$ gives the intensity distribution, and $S(x,z) = (1/2i) \ln \left[ \psi_0(x,z)/\psi_0^*(x,z) \right]$ is the associated phase field.
Accordingly, in order to obtain global information about the SPP propagation is the transversal intensity, defined as
\begin{equation}
	\bar{I}_0(z) = \int I_0(x,z) dx .
	\label{eq8}
\end{equation}
The substituting of Eq.~\eqref{eq7} into \eqref{eq8} leads to
\begin{equation}
	\bar{I}_0(z) = \frac{2\pi}{|k_{\rm SPP}|} \int |\tilde{\psi}_0(u)|^2 e^{-k''_{\rm SPP} z u^2} du .
	\label{eq9}
\end{equation}
Furthermore, using Eq.~\eqref{eq6}, we can readily find that
\begin{equation}
	\frac{d\bar{I}_0(z)}{dz} = - \frac{k''_{\rm SPP}}{|k_{\rm SPP}|^2} \int \left\arrowvert \frac{\partial \psi_0(x,z)}{\partial x}\right\arrowvert^2 dx .
	\label{eq10}
\end{equation}
If we now appeal againg to Eq.~\eqref{eq7} and substitute it into \eqref{eq10}, then we obtain
\begin{equation}
	\bar{I}'_0(z) := \frac{d\bar{I}_0(z)}{dz} = - \frac{2\pi k''_{\rm SPP}}{|k_{\rm SPP}|} \int u^2 |\tilde{\psi}_0(u)|^2 e^{-k''_{\rm SPP} z u^2} du .
	\label{eq11}
\end{equation}
Dividing Eq.~\eqref{eq11} by \eqref{eq9} finally leads to a more compact form to describe how the transversal intensity changes along the $z$-direction,
\begin{equation}
	\frac{1}{\bar{I}_0(z)} \frac{d\bar{I}_0(z)}{dz} = - k''_{\rm SPP} \mu_2 (z) ,
	\label{eq12}
\end{equation}
where
\begin{equation}
	\mu_2 (z) = \frac{\displaystyle \int u^2 |\tilde{\psi}_0(u)|^2 e^{-k''_{\rm SPP} z u^2} du}{\displaystyle \int |\tilde{\psi}_0(u)|^2 e^{-k''_{\rm SPP} z u^2} du} .
	\label{eq13}
\end{equation}
According to the above findings, we note that the transversal intensity is going to depend on the second-order moment of $|\tilde{\psi}_0(u)|^2 e^{-k''_{\rm SPP} z u^2}$, thus being independent of the phase of $\tilde{\psi}_0(u)$, which is rather convenient to compare the properties of different types of bending SPP (Airy, Pearcy, Olver, Butterfly, etc.).
It is also worth mentioning that, in those cases where $k''_{\rm SPP} = 0$, the second-order moment becomes independent of $z$, as it can readily be seen in Eq.~\eqref{eq11}.
This term is usually referred to as beam divergence and is related to the beam quality parameter \cite{serna:JOSAA:1991,siegman:SPIE:1993}.

We respect to the phase of $\psi_0(x,z)$, we can proceed in a similar manner and compute the average phase variations along the $z$-direction,
\begin{equation}
	\left\langle \frac{\partial S}{\partial z} \right\rangle\! (z) = \int I_0(x,z) \frac{\partial S(x,z)}{\partial z} dx ,
	\label{eq14b}
\end{equation}
which leads to the following relation with the transversal intensity
\begin{equation}
	\left\langle \frac{\partial S}{\partial z} \right\rangle\! (z) = \frac{k'_{\rm SPP}}{2 k''_{\rm SPP}} \frac{d\bar{I}_0(z)}{dz} = - \frac{k'_{\rm SPP}}{2}\ \bar{I}_0(z) \mu_2 (z) .
	\label{eq15}
\end{equation}
These relations show that it suffices to know the evolution of the transversal intensity to determine the average phase derivative of a SPP and vice versa, this being a feature that could be profitably exploited at the experimental level to retrieve one of the two quantities by measuring the other one.

%%%%%%%%%%%%%%%%%%%%%%%%%%%%%%%%%%%%%%%%%%%%%%%%%%%%%%%%%%%%%%%%%%%%%%%%
%%%%%%%%%%%%%%%%%%%%%%%%%%%%%%%%%%%%%%%%%%%%%%%%%%%%%%%%%%%%%%%%%%%%%%%%

\section*{Acknowledgments}
This research has been partially supported by Grants No.~PID2021-127781NB-I00 (ASS) and  No.~PID2022-137569NB-C42 (RMH) both funded by MCIN/AEI/10.13039/501100011033 and the ``European Union NextGenerationEU/PRTR'' (RMH); the FASLIGHT Network, Grant No.~RED2022-134391-T (RMH), funded by MCIN/AEI/10.13039/501100011033; and by Project 2020-T1/IND-19951 (JHR) funded by the ``Programa de Atracci\'on de Talento de la Comunidad Aut\'onoma de Madrid (Modalidad~1)''.

%%%%%%%%%%%%%%%%%%%%%%%%%%%%%%%%%%%%%%%%%%%%%%%%%%%%%%%%%%%%%%%%%%%%%%%%
%%%%%%%%%%%%%%%%%%%%%%%%%%%%%%%%%%%%%%%%%%%%%%%%%%%%%%%%%%%%%%%%%%%%%%%%

  \bibliographystyle{elsarticle-num} 
%  \bibliography{references}
%  \biboptions{sort&compress}


%\end{document}

\biboptions{sort&compress}

\end{document}